\documentclass[aps,preprintnumbers,amsmath,amssymb,floatfix,nofootinbib]{revtex4}
\usepackage{epsf}
\usepackage{graphicx}
\usepackage{subfigure}
\usepackage{color}
\newcommand{\be}{\begin{equation}}
\newcommand{\ee}{\end{equation}}
\newcommand{\ba}{\begin{eqnarray}}
\newcommand{\ea}{\end{eqnarray}}

\newcommand{\grts}{\raise.3ex\hbox{$>$\kern-.75em\lower1ex\hbox{$\sim$}}}
\newcommand{\lets}{\raise.3ex\hbox{$<$\kern-.75em\lower1ex\hbox{$\sim$}}}

\begin{document}
\title{Projections for Two Higgs Doublet Models at the LHC and ILC: \\A Snowmass White Paper}
\author{Chien-Yi Chen}

\affiliation{
Department of Physics, Brookhaven National Laboratory, Upton, New York, 11973}

\date{\today}

\begin{abstract}
The discovery of the 125 GeV Higgs boson and the measurement of its branching ratios has initiated the 
exploration of the electroweak symmetry breaking sector.   
There have been numerous studies exploring the restrictions these results place on the parameter space of two Higgs doublet models.  
 We extend these results  to include the full data set and study the expected sensitivity that can be obtained with 
  $300~$fb$^{-1}$ and $3000$~fb$^{-1}$ integrated luminosity.
In addition, searches for a heavy Standard Model Higgs boson are also considered. It is shown that the nonobservation of 
such a Higgs boson can substantially narrow the allowed regions of parameter space in two Higgs doublet models. 
Finally, projections for the ILC at the center of mass energy 250, 500, and 1000 GeV are investigated. 
\end{abstract}

\maketitle

\begin{table}[t]
\caption{Light Neutral Higgs ($h^0$) Couplings in the 2HDMs}
\centering
\begin{tabular}[t]{|c|c|c|c|c|}
\hline\hline
& I& II& Lepton Specific& Flipped\\
\hline
$g_{hVV}$ & $\sin(\beta-\alpha)$ & $\sin (\beta-\alpha)$ &$\sin (\beta-\alpha)$&$\sin (\beta-\alpha)$\\
$g_{ht\overline{t}}$&${\cos\alpha\over\sin\beta}$&${\cos\alpha\over\sin\beta}$&${\cos\alpha\over\sin\beta}$&${\cos\alpha\over\sin\beta}$\\
$g_{hb{\overline b}}$ &${\cos\alpha\over\sin\beta}$&$-{\sin\alpha\over\cos\beta}$&${\cos\alpha\over\sin\beta}$&$-{\sin\alpha\over \cos\beta}$\\
$g_{h\tau^+\tau^-}$&${\cos\alpha\over \sin\beta}$&$-{\sin\alpha\over \cos\beta}$&$-{\sin\alpha\over \cos\beta}$&${\cos\alpha\over\sin\beta}$\\
\hline
\end{tabular}
\label{table:coups}
\end{table}

\begin{table}[t]
\caption{Heavy Neutral CP Even Higgs ($H^0$) Couplings in the 2HDMs}
\centering
\begin{tabular}[t]{|c|c|c|c|c|}
\hline\hline
& I& II& Lepton Specific& Flipped\\
\hline
$g_{HVV}$ & $\cos(\beta-\alpha)$ & $\cos (\beta-\alpha)$ &$\cos (\beta-\alpha)$&$\cos (\beta-\alpha)$\\
$g_{Ht\overline{t}}$&${\sin\alpha\over\sin\beta}$&${\sin\alpha\over\sin\beta}$&${\sin\alpha\over\sin\beta}$&${\sin\alpha\over\sin\beta}$\\
$g_{Hb{\overline b}}$ &${\sin\alpha\over\sin\beta}$&${\cos\alpha\over\cos\beta}$&${\sin\alpha\over\sin\beta}$&${\cos\alpha\over \cos\beta}$\\
$g_{H\tau^+\tau^-}$&${\sin\alpha\over \sin\beta}$&${\cos\alpha\over \cos\beta}$&${\cos\alpha\over \cos\beta}$&${\sin\alpha\over\sin\beta}$\\
\hline
\end{tabular}
\label{table:coupsH}
\end{table}

\section{Introduction}
The discovery of a Higgs boson at the LHC with a mass around 125 GeV is the beginning of the exploration of the source of electroweak symmetry breaking.
Many beyond the Standard Model (SM) scenarios have a Higgs-like particle which has SM-like couplings to fermions and gauge bosons. 
A well-motivated extension of the SM is obtained by adding a second SU(2)$_L$ Higgs doublet, leading to
five physical Higgs scalars: two charged Higgs bosons $H^\pm$, a pseudoscalar $A$,
and two neutral scalars, $h^0$ and $H^0$. Although it is possible that the $125$ GeV state is the heavier neutral Higgs particle,  
$h^0$,  of a $2$ Higgs doublet model (2HDM)
\cite{Ferreira:2012my,Drozd:2012vf,Chang:2012ve}, we assume here that it is the lighter. 
Two Higgs doublet models generically have tree level flavor changing neutral
currents from Higgs exchanges unless there is a global or discrete symmetry which forbids such
 interactions\cite{Branco:2011iw,Grossman:1994jb} and therefore we consider only
the class of models where there is a discrete $Z_2$ symmetry such that one type of the fermions couples only to a single Higgs doublet.
There are four possibilities for 2HDMs of this type which are
typically called the type-I, type-II, lepton specific, and flipped models\cite{Branco:2011iw}.
The couplings of the Higgs bosons to fermions are described by two free parameters.   
The ratio of vacuum expectation values of the two Higgs doublets is  $\tan\beta \equiv \frac{v_2}{v_1}$, 
and the mixing angle which diagonalizes the neutral scalar mass matrix is $\alpha$.
The couplings of a neutral CP-even Higgs ($\phi^0$) to the SM particles are parametrized as
 %fermions ($g_{ii\phi}$) and vector bosons ($g_{VV\phi}$) are parametrized as
%leading to 5 physical Higgs particles: 
%$h^0, H^0,A^0$, and $H^\pm$. In this paper, I will assume that the observed Higgs boson is the lightest one, $h^0$. 
\begin{equation}
{\cal L}=-\sum_f g_{\phi f\bar{f}} {m_f\over v} {\overline f} f\phi^0 -\sum_{V=W,Z} g_{\phi VV}{2M_V^2\over v} V_\mu V^\mu \phi^0\, ,
\end{equation}
where $\phi^0$ can be $h^0$ or $H^0$. 
The couplings of $\phi^0$ to the fermions and vector bosons are $g_{\phi f\bar{f}}$ and $g_{\phi VV}$, respectively.
$m_f$ and $M_V$ are masses of the fermions and vector bosons, respectively.
Moreover, $g_{hf\bar{f}}=g_{hVV}=1$ and $g_{Hf\bar{f}}=g_{HVV}=0$ in the SM and $v=\sqrt{v_1^2+v_2^2}\sim246$ GeV.
The $\phi^0$ coupling to gauge bosons is the same for all four models considered here, while the couplings to fermions 
differentiates between the models. 
The couplings of $h^0$ and $H^0$ to fermions and gauge bosons
relative to the SM couplings
 are given for all four 2HDMs in Table \ref{table:coups} and Table \ref{table:coupsH}, respectively. 
%Many papers Ref. \cite{Barger:2013mga,Barger:2013ofa} have studied the 
Many papers \cite{Ferreira:2011aa,Chen:2013kt,Chen:2013rba,Alves:2012ez,Craig:2012vn,Craig:2012pu,Bai:2012ex,Azatov:2012qz,Dobrescu:2012td,
Ferreira:2012nv,Celis:2013rcs,Grinstein:2013npa,Shu:2013uua,Krawczyk:2013gia,Coleppa:2013dya,Altmannshofer:2012ar,
Chiang:2013ixa,Basso:2012st,Barroso:2013zxa,Barroso:2013awa,Chang:2012zf,Craig:2013hca,Barger:2013mga,Barger:2013ofa,Lopez-Val:2013yba,
Ferreira:2013qua,Freitas:2012kw,Kanemura:2013eja,Belanger:2013xza,Rindani:2013mqa,Harlander:2013mla} 
examined various channels in the four 2HDMs in light of the experimental findings at the LHC and ILC. In this paper,
we study constraints on the parameter space in 2HDMs using 
current Higgs coupling measurements as well as heavy Higgs searches at the LHC, the extension of these limits to 300 fb$^{-1}$ and 3000 fb$^{-1}$,
and projections for the ILC at the center of mass energy ($\sqrt{s}$) 250, 500 and 1000 GeV.

\section{LHC Reach from $h^0$ Measurements}

We have shown the results with the most recent experimental
data and have also studied the bounds that can be obtained at a future LHC with 14 TeV and 
 integrated luminosities of 300 fb$^{-1}$ and
3000 fb$^{-1}$.  To estimate these bounds, we look at the current errors, assume that the SM 
prediction is correct, and scale the errors as $1/\sqrt{N}$, where $N$ scales like the integrated luminosity.   
This corresponds to `scheme 2' of the CMS\cite{nisati} high luminosity projections.

A $\chi^2$ fit to the data shown in Tables \ref{tab:models1} and \ref{tab:models2}  is performed assuming 
$M_{h^0}=125$ GeV. We follow the standard definition of $\chi^2 = \sum_i {(R_i^{\rm 2HDM}-R_i^{\rm meas})^2\over (\sigma^{meas}_i)^2}$, where $R^{2HDM}$ 
represents predictions for the signal strength from the 2HDMs and $R^{\rm meas}$ stands for the measured 
signal strength shown in Tables \ref{tab:models1} and \ref{tab:models2}.

\begin{figure}[tb]
\subfigure[]{
      \includegraphics[width=0.36\textwidth,angle=0,clip]{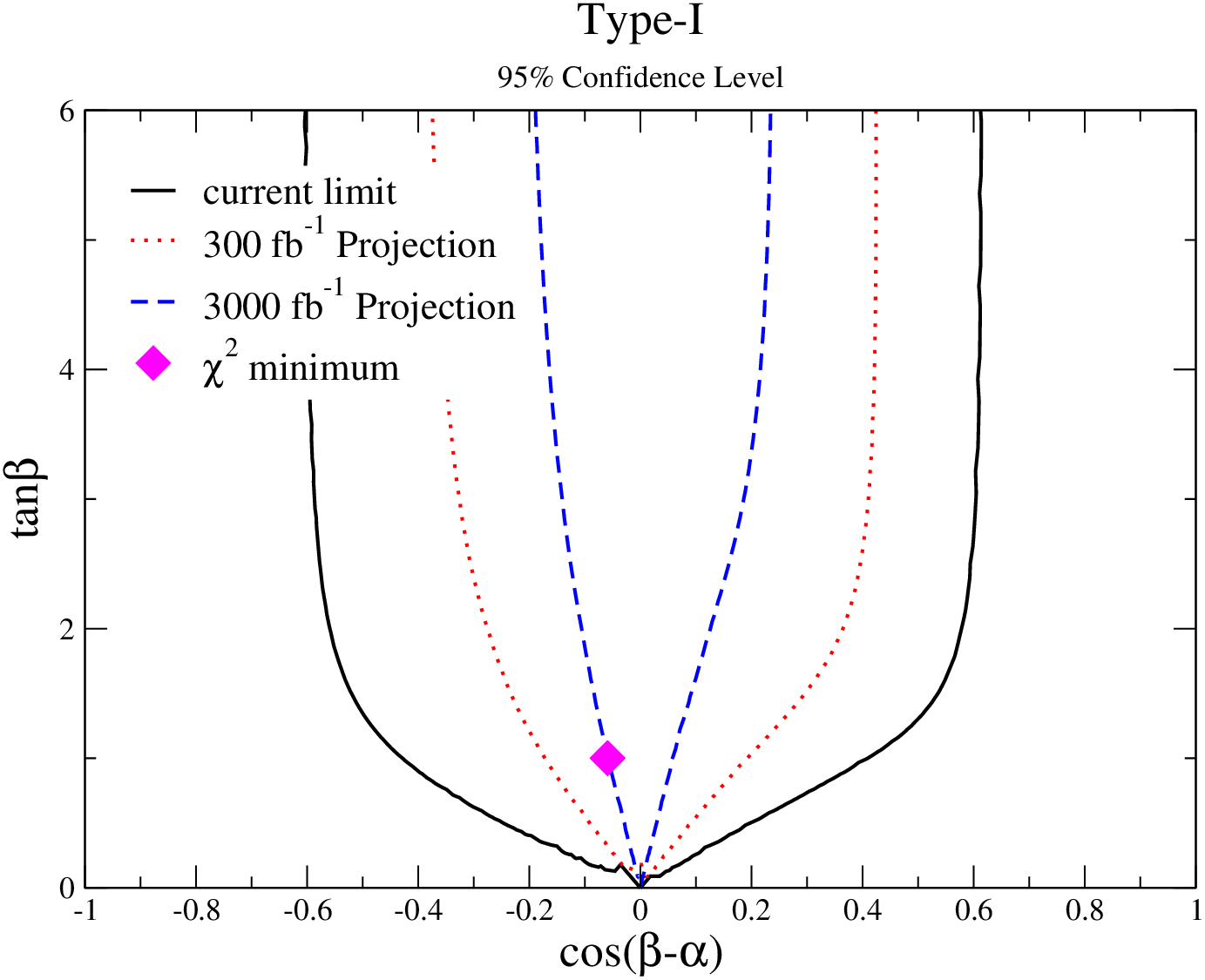}
}
\subfigure[]{
      \includegraphics[width=0.36\textwidth,angle=0,clip]{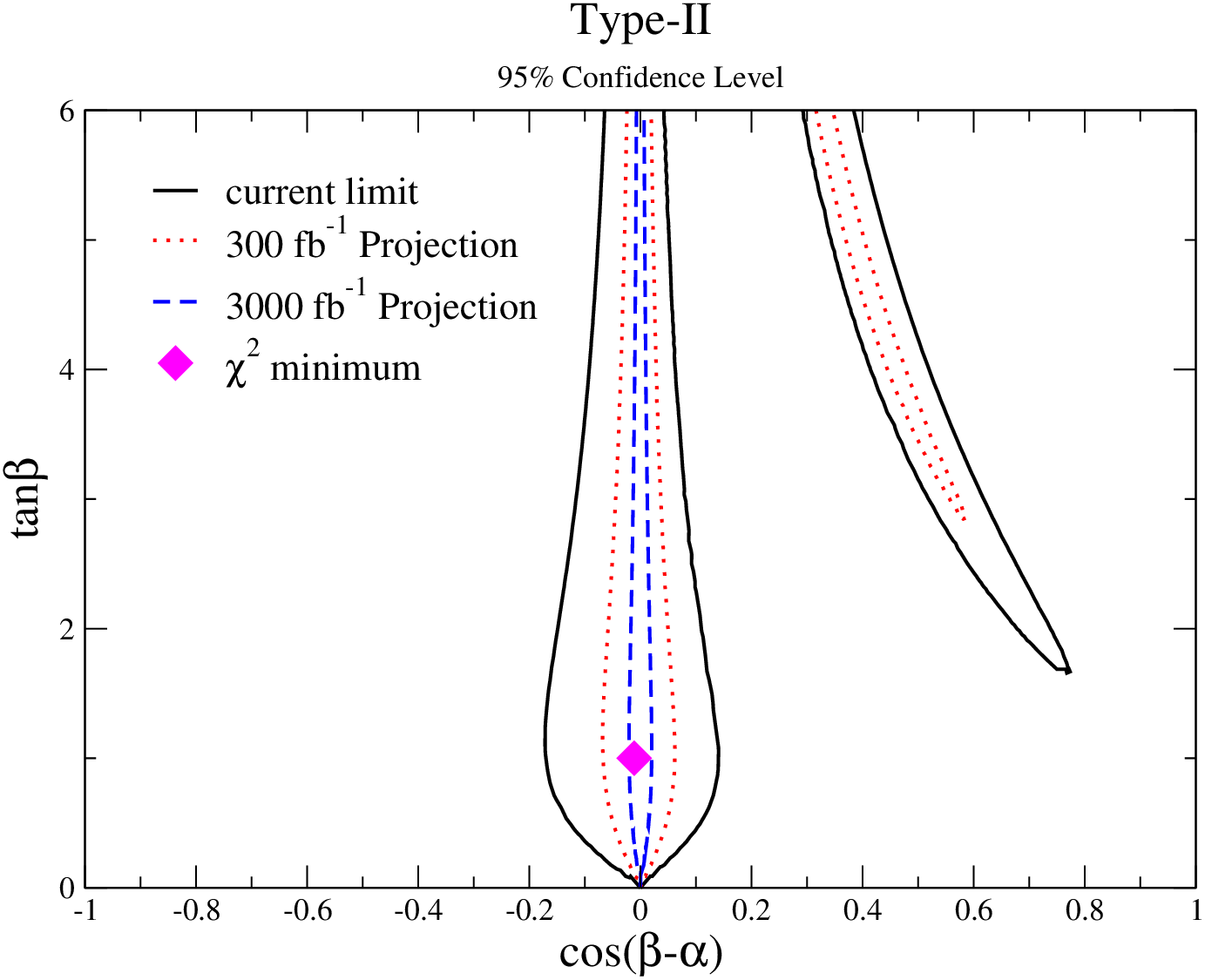}
}
\subfigure[]{
      \includegraphics[width=0.36\textwidth,angle=0,clip]{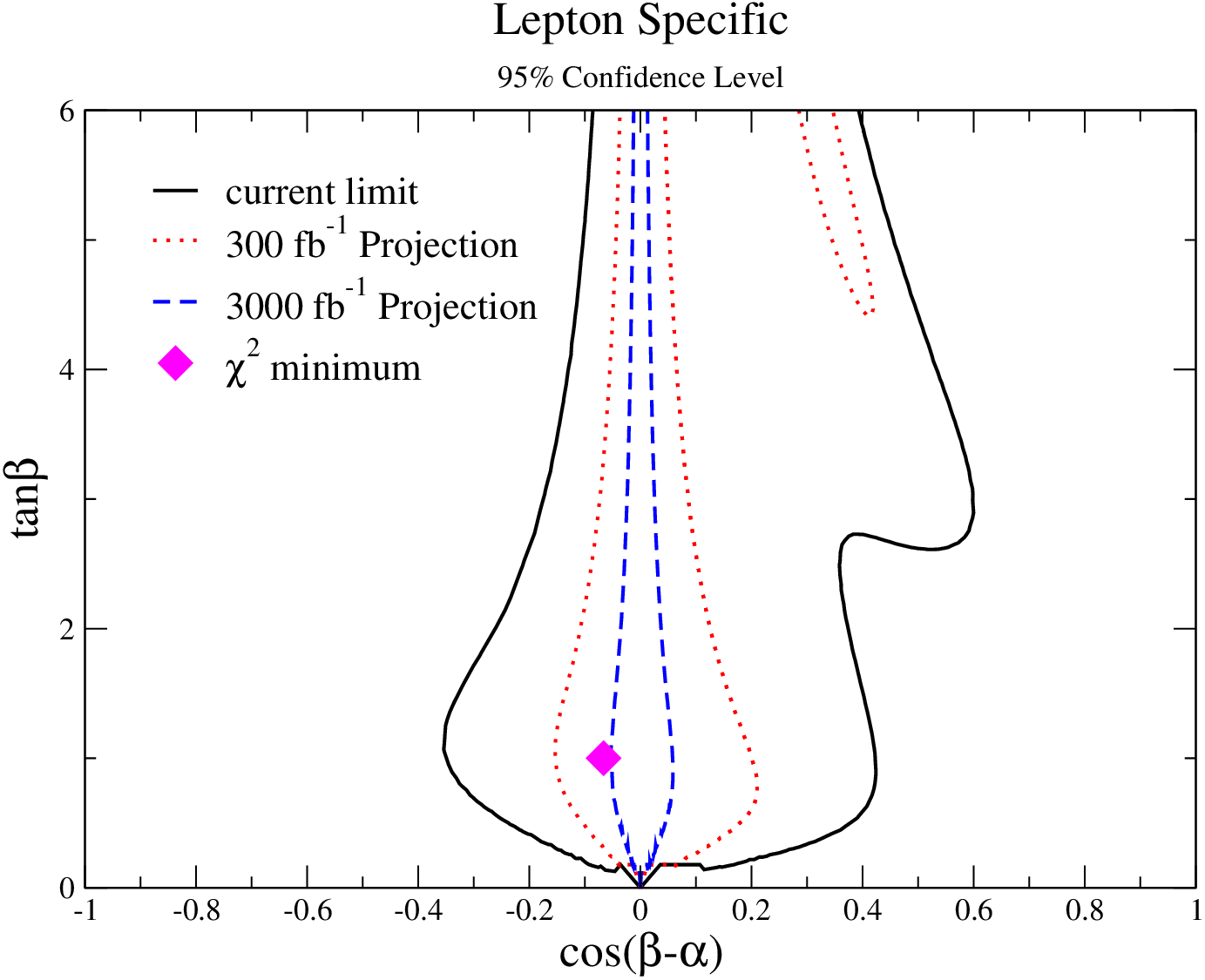}
}
\subfigure[]{
      \includegraphics[width=0.36\textwidth,angle=0,clip]{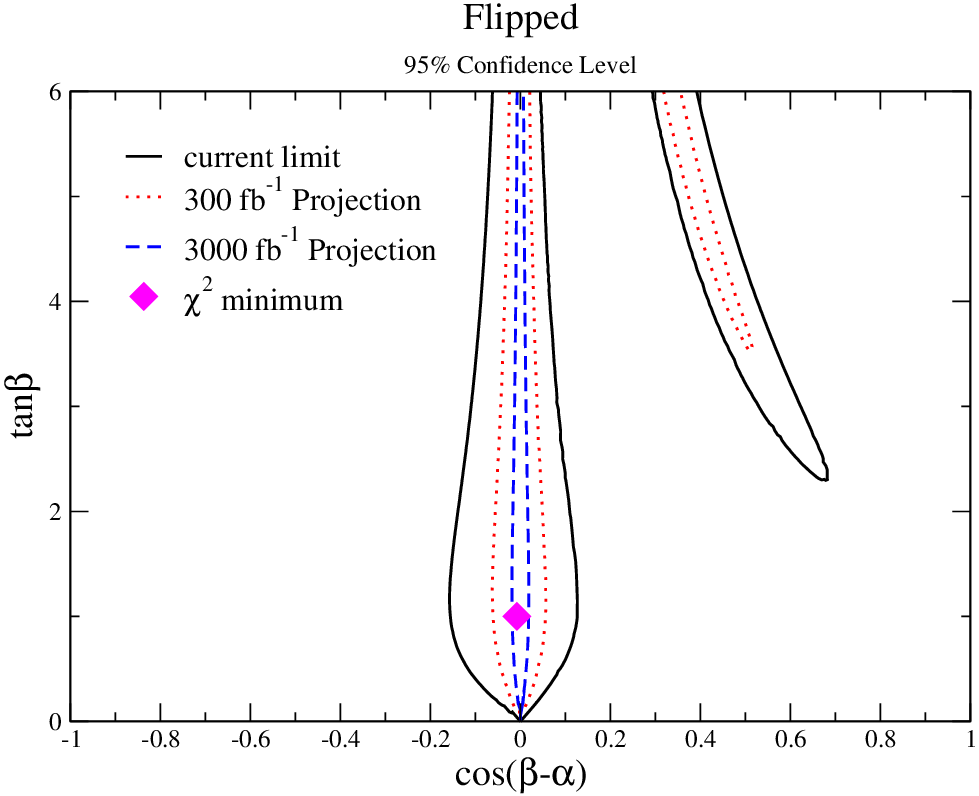}
}
\caption{Allowed regions  in the $(\cos(\beta-\alpha),\tan\beta)$ plane  in type-I (a), type-II (b), Lepton Specific (c),
and Flipped (d) 2HDMs obtained by performing  a $\chi^2$ analysis.  
The region between the black (solid), red (dotted), and blue (dashed)
 lines is allowed at $95\%$ confidence level corresponding to the current limits and the projected limits for integrated luminosities of 
 300 fb$^{-1}$ and 3000 fb$^{-1}$, respectively.
}
\label{chisq_fig}
\end{figure}

Our results are given in Fig. \ref{chisq_fig}.
\begin{itemize}
 \item  For each of the four models, we plot the current limits on the parameter space, and the
 projected limits for integrated luminosities of $300~$fb$^{-1}$ and $3000~$fb$^{-1}$. 
 Bounds from flavor physics  constrain $\tan\beta \ge1$ \cite{Chen:2013kt,Mahmoudi:2009zx}
 and we take this as a prior when we determine the chi-squared minima.
\item In all of the models the minimum of the $\chi^2$ occurs for $\tan\beta\sim 1$ and $\cos(\beta-\alpha)\sim 0$, 
demonstrating that the couplings of a 2HDM are already constrained
to be close to the SM values.   
\item The parameter space for the type-I model is not very constrained at present.  This is because, in the large $\tan\beta$ 
 limit, %the Higgs is fermiophobic and production through gluon fusion is suppressed.
 $g_{hf \bar{f}} = { \cos\alpha\over\sin\beta} \approx {\cos\alpha}$,  which is independent of tan$\beta$. 
 Increasing the integrated luminosity will gradually narrowed the allowed parameter space in the large $\tan\beta$ 
 limit, 
\item The lepton specific model is also not severely constrained, because of the enhanced 
decay to $\tau$ leptons, which is poorly
 measured at present.  For large $\tan\beta$, the bottom-quark Yukawa coupling becomes substantial 
 in the type-II and flipped models, and thus the 
 currently allowed parameter space is much more restricted.   
\end{itemize}

\section{Constraints from Heavy Higgs Searches}
ATLAS \cite{Aad:2012tfa} and CMS \cite{Chatrchyan:2013yoa} have obtained  upper bounds on the ratio  of the production cross section to the SM expectation for a SM Higgs boson 
in various decay channels with a mass between 150 and 600 GeV and assuming a SM width.   
 We use the $95\%$ confidence level band from recent CMS bounds (from Figure 11 in Ref.\cite{Chatrchyan:2013yoa})
 and  scale predictions as the inverse square root of the integrated luminosity.
For example, suppose $M_{H^0}$ is 200 GeV.   A SM Higgs boson
 of $200$ GeV will decay almost $100\%$ of the time into vector bosons.  
This is also true (except for extreme values of the parameters) in a 2HDM.    
 The production rate through gluon fusion in 2HDM will be different than the SM rate
  because of the different $t$ and $b$ couplings.
Thus, the upper bound from ATLAS and CMS on the cross section relative to the SM rate will  
   place a constraint on $\alpha$ and $\beta$. 

\begin{figure}[tb]
\subfigure[]{
      \includegraphics[width=0.36\textwidth,angle=0,clip]{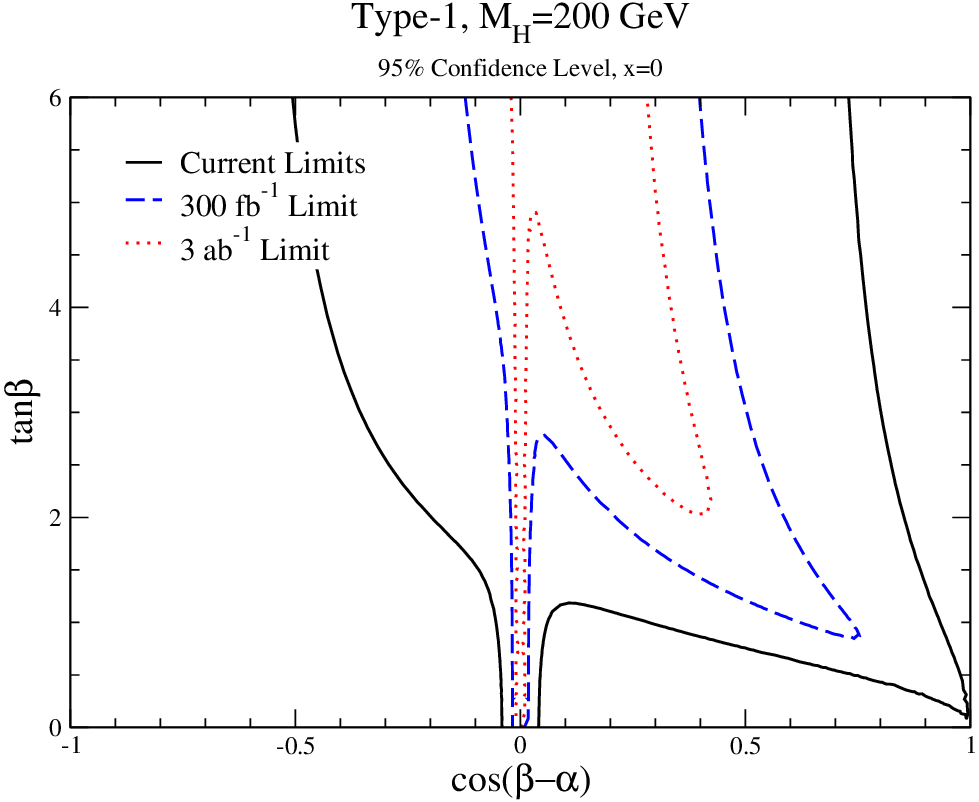}
}
\subfigure[]{
      \includegraphics[width=0.36\textwidth,angle=0,clip]{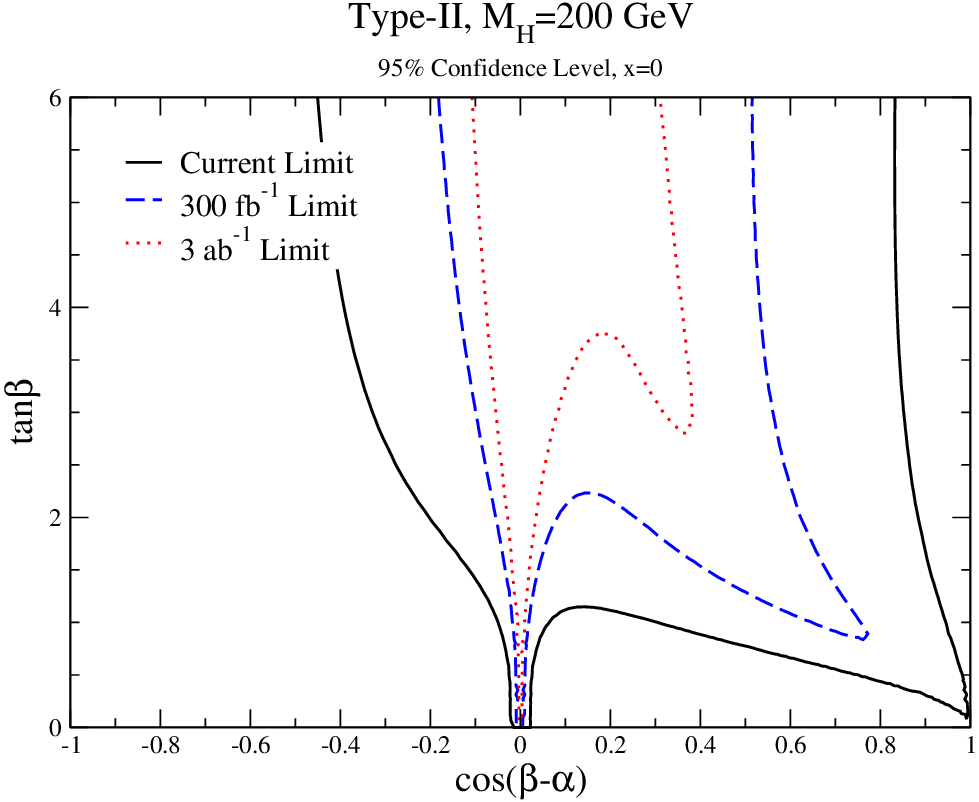}
}
\caption{ Allowed regions in type-I (a) and type-II (b) 2HDMs from the LHC limit on a 200 GeV heavy Higgs boson.
The region between the black (solid), blue (dashed) and red (dotted) curves is allowed at 95\% confidence level corresponding to
the current limits and the projected limits for integrated luminosities of 300 fb$^{-1}$ and 3000 fb$^{-1}$, respectively.
}
\label{fg:mh2}
\end{figure}
We find that the best constraint comes from $M_{H^0} = 200$ GeV, and therefore we will just show the result at this mass point.
More detailed analyses can be found in Ref.\cite{Chen:2013rba}.
For $M_{H^0} = 200$ GeV, the results are summarized in Fig. \ref{fg:mh2}.
    We show results for the type-I and type-II models, with the current limits and projections for
    300 fb$^{-1}$ and 3000 fb$^{-1}$.   
\begin{itemize}
 \item   
    The lepton specific and flipped models give very similar results to the type-I and type-II models, respectively. 
    An increase in luminosity will tightly constrain $\cos(\beta-\alpha)$ for $\tan\beta < 4$ in the type-I model 
    and will give a significant constraint for $\tan\beta < 4$ in the type-II model.  
\item    
We see that even with current bounds,
 a significant fraction of the previously allowed parameter space in the type-I model is excluded by the heavy Higgs search results, and this
  fraction  grows with increasing integrated luminosity (unless, of course, the heavy Higgs is discovered).     
  For the type-II model, some of the remaining parameter space is excluded, especially for small $\tan\beta$.   
  This is a significant result, and shows that the allowed parameter space of a 2HDM can be substantially narrowed by considering bounds 
  from heavy Higgs searches.
\end{itemize}

\section{Constraints from the ILC}
\begin{figure}[tb]
\subfigure[]{
      \includegraphics[width=0.36\textwidth,angle=0,clip]{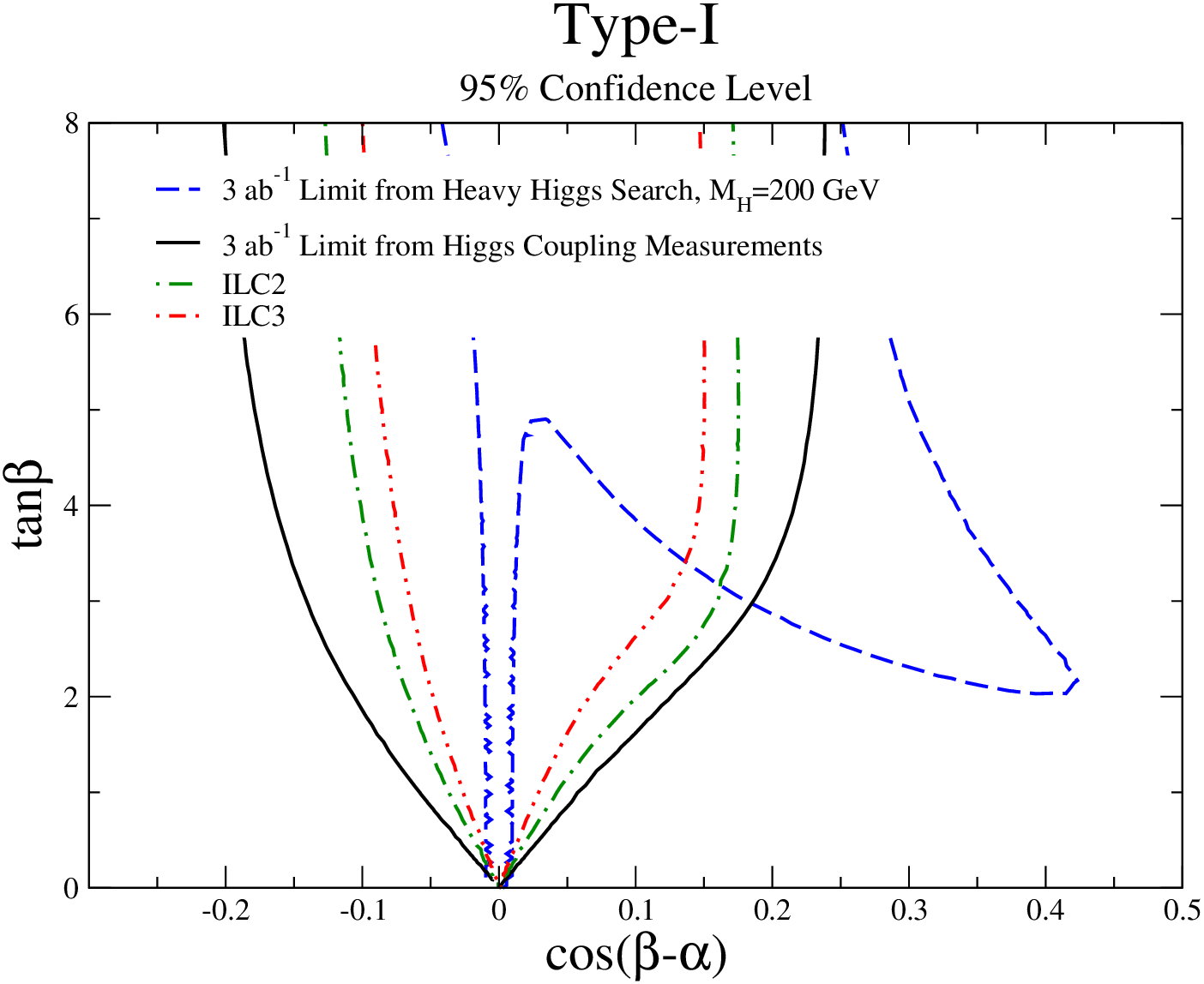}
}
\subfigure[]{
      \includegraphics[width=0.36\textwidth,angle=0,clip]{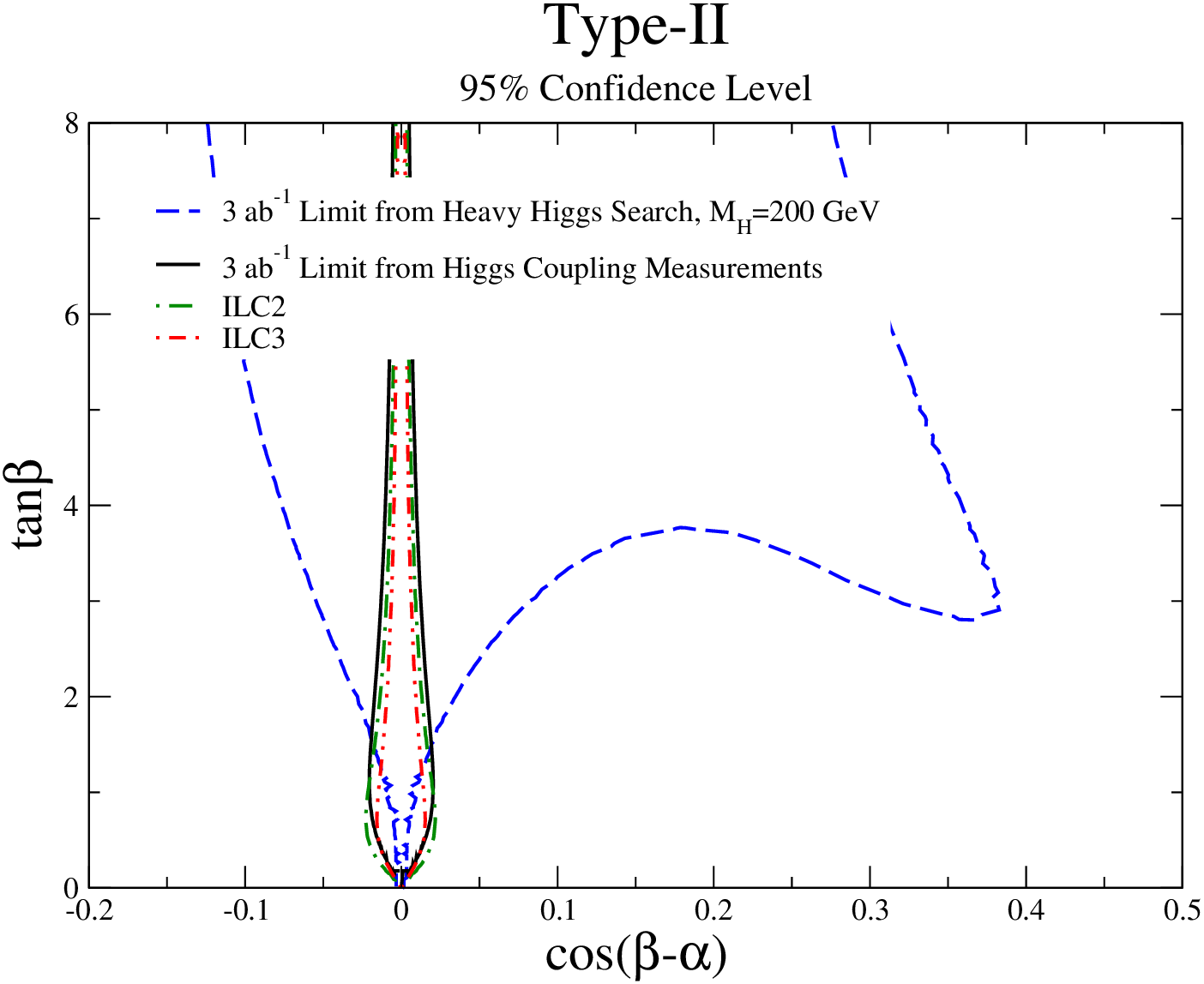}
}
\subfigure[]{
      \includegraphics[width=0.36\textwidth,angle=0,clip]{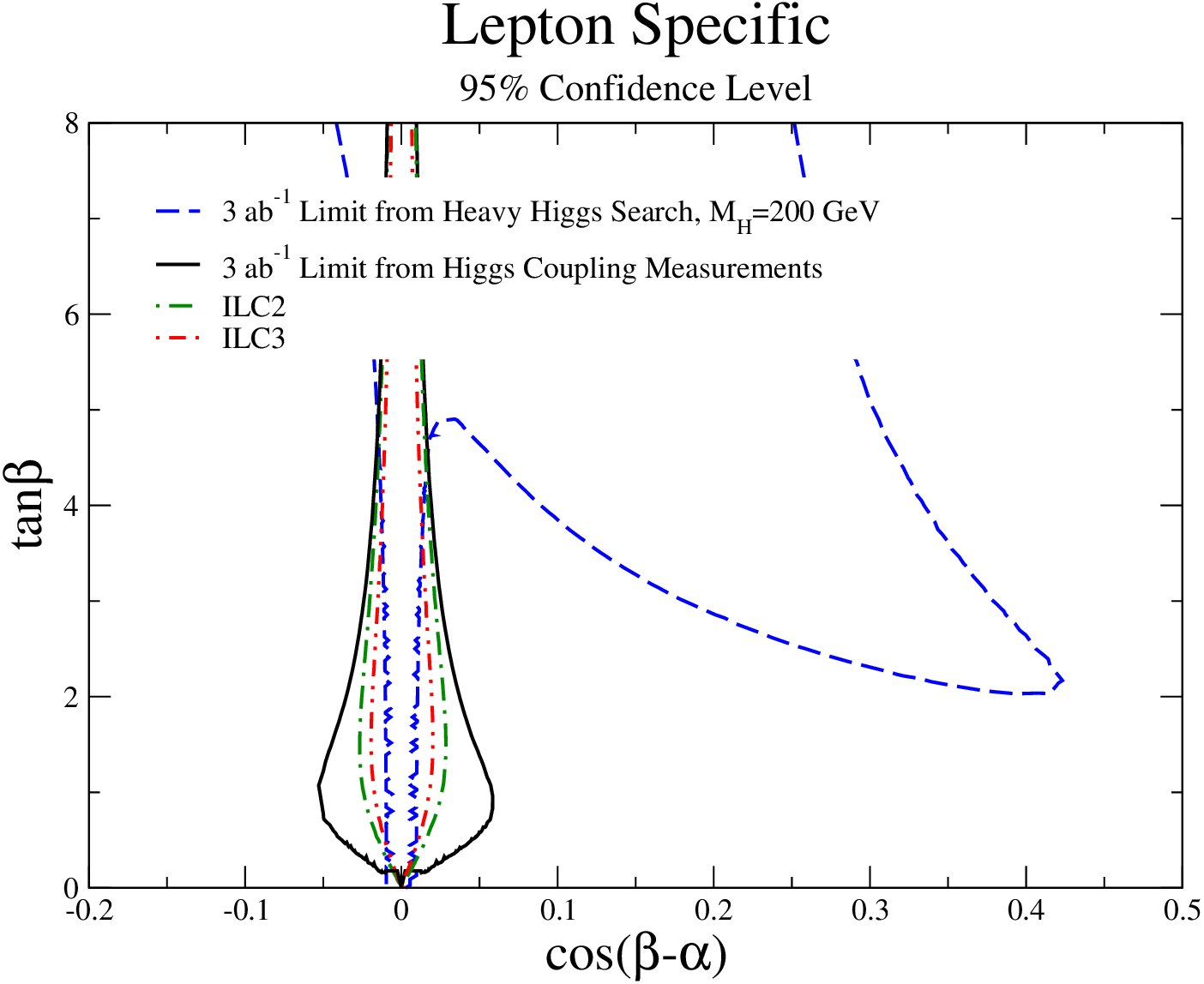}
}
\subfigure[]{
      \includegraphics[width=0.36\textwidth,angle=0,clip]{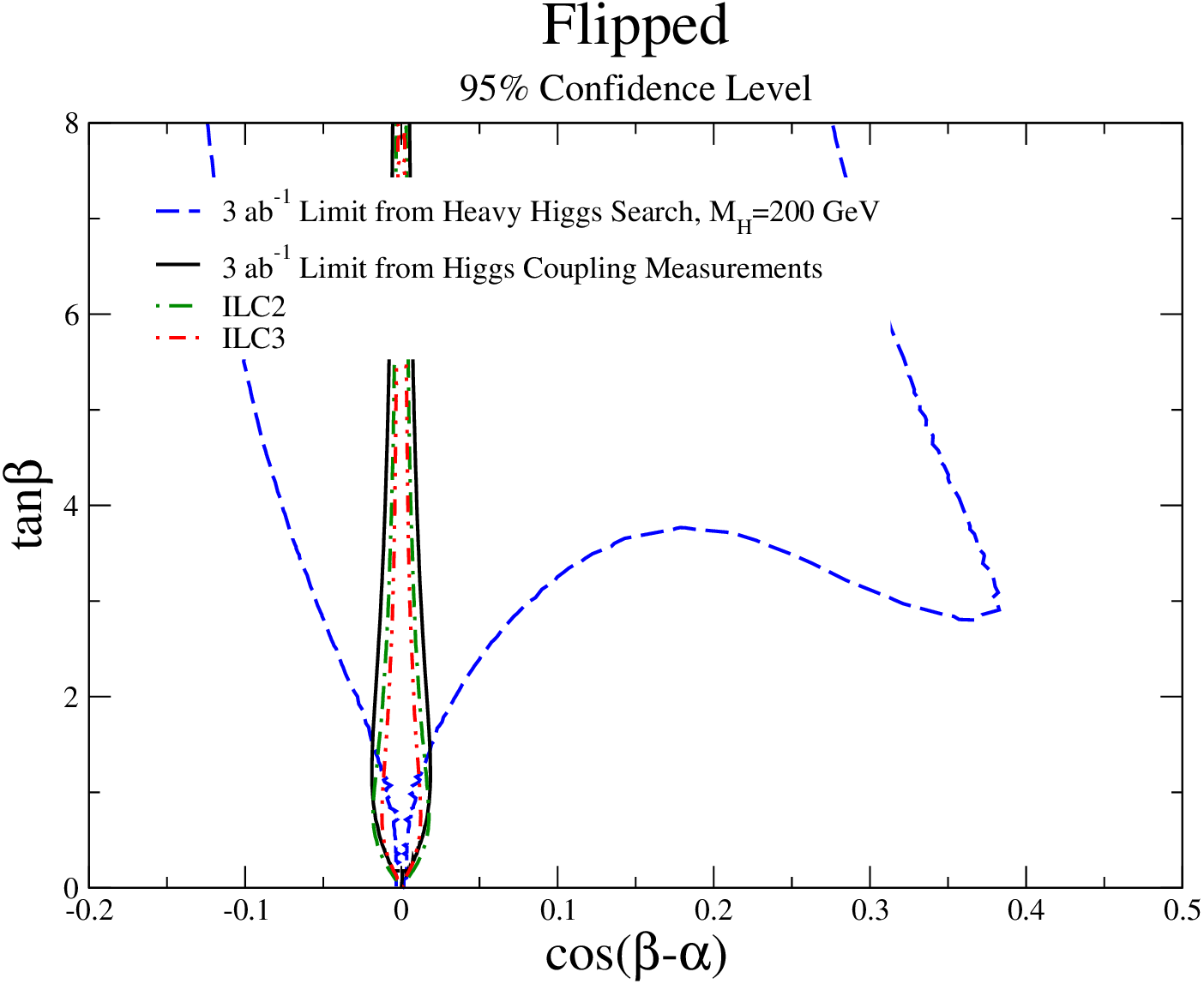}
}
\caption{Allowed regions  in the $(\cos(\beta-\alpha),\tan\beta)$ plane  in type-I (a), type-II (b), lepton specific (c),
and flipped (d) 2HDMs obtained by performing a $\chi^2$ analysis.  
The region between lines is allowed at $95\%$ confidence level. 
The black (solid) lines represent the LHC coupling measurements at 3 ab$^{-1}$. The blue (dashed) lines stand for bounds from heavy 
Higgs search at 3 ab$^{-1}$. The green (dot-dashed) and red (double dot-dashed) lines correspond to the projections 
of ILC2 and ILC3, respectively.
 %corresponding to the current limits and the projected limits for integrated luminosities of 
 %$300~fb^{-1}$ and $3000~ fb^{-1}$, respectively.
}
\label{fg:ilc}
\end{figure}

The ILC has great potential to precisely measure the total width and mass of the discovered Higgs. 
It is important to study the projected sensitivity at the ILC and its constraints on the parameter space of 2HDMs.
The design center of mass energy at the ILC are 250 and 500 GeV with a possibility to upgrade to 1 TeV. 
The corresponding integrated luminosities are
250, 500 and 1000 fb$^{-1}$, respectively. In the following, we will use the notations:
\begin{itemize}
 \item ILC1: data collected from ILC at $\sqrt{s}$ = 250 GeV with a luminosity of 250 fb$^{-1}$.
 \item ILC2: ILC1 plus data collected from ILC at $\sqrt{s}$ = 500 GeV with a luminosity of 500 fb$^{-1}$.
 \item ILC3: ILC2 plus data collected from ILC at $\sqrt{s}$ = 1000 GeV with a luminosity of 1000 fb$^{-1}$.
\end{itemize}
The uncertainties used in our $\chi^2$ analysis are taken from Table 2 of Ref.\cite{Peskin}, which can also be found
in Ref.\cite{Adolphsen:2013jya}. 
%We use cumulative luminosities in order to take the advantage of using the full data set. 
%In other words, for ILC3, all the data with lower center of mass energy and luminosities are also considered.  
%Here we make a comparison of ILC and LHC future projection.
It is very interesting to compare future projections for the LHC and ILC, as shown in Fig. \ref{fg:ilc}.
\begin{itemize}
 \item 
In the type-I model, 
%large part of parameter space has already been excluded at ILC2. 
ILC2 already improves the limit a lot compared with that from LHC light higgs 
measurements. Limits from ILC3 further restrict the parameter space a bit.
%It is clear that an increase in  luminosity at the ILC narrows down the allowed parameter space. 
Furthermore, bounds from the heavy Higgs search are very restrictive at low tan$\beta$ (tan$\beta<4$). 
%This is due to the weak limits occur at cos($\beta-\alpha$)=0 
%
 \item
In the type-II model, only a small region of the parameter space is allowed at 95\% CL. 
For the large tan$\beta$ region, limits from the heavy Higgs search at the LHC are weaker compared with those from
the coupling measurements at the LHC and ILC, which tightly constrain the parameter space. 
At low tan$\beta$, however, constraints from the heavy Higgs searches strongly restrict the parameter space.
%insert A
%Limits from ILC3 can improve the bounds a bit at low tan$\beta$.
%
 \item
For the lepton specific model, all bounds except for those from the LHC heavy Higgs search are similar in the large tan$\beta$ region,
but bounds from the ILC look slightly restrictive.
ILC2 and ILC3 are able to give better limits than those from the LHC light Higgs measurements for tan$\beta <$ 4.
%
%Going from ILC500 to ILC1TeV shrink the parameter space a bit.
 \item
In the type-II model, the allowed region centers around cos($\beta-\alpha$)=1 and becomes narrower as tan$\beta$ increases. 
This is because in this model the Higgs coupling
to the b quarks, $g_{hb \bar{b}} = { \sin\alpha\over\cos\beta}$, is proportional to $\sqrt{1+\tan^2\beta}$. The bound gets stronger 
when tan$\beta$
becomes larger. This is also true for the flipped model. As a result, bounds on the parameter space of the type-II and the flipped models 
are very similar.
%Likewise, for the lepton specific model, $g_{h\tau\bar{\tau}}  \propto$ tan$\beta$ so the same argument is applicable to 
%this model.
\end{itemize}
%=====================================================================================comparison ILC with LHC

\section{Conclusions}
In this paper, we examined the projected restrictions on 2HDM parameter space from the LHC with 300 and 3000 fb$^{-1}$ of data.
In particular, we focused on the sensitivity obtained via measurements of the 125 GeV Higgs boson production and decay rates, 
and demonstrated that LHC searches for SM Higgs bosons can be recast as searches for other heavy Higgs bosons of 2HDMs.
For the type-I model with a heavy Higgs mass of 200 GeV, the parameter space allowed from 
  branching ratios of the 125 GeV Higgs can
   be shrunk by more than a factor of two by including bounds from the heavy Higgs searches.   
   It is thus important, in the LHC upgrade, to continue these searches.
For all models bounds from heavy Higgs searches are very constraining at low tan $\beta$ (tan$\beta<$4), while they are
less restrictive at larger tan$\beta$. However, bounds from the coupling measurements at the LHC and ILC can 
constrain the large tan$\beta$ region very well but are less constraining at low tan$\beta$. 
In particular, for the type-I model, the ILC can give more restrictive bounds at large tan$\beta$.
As a consequence, we find that searches at the LHC and ILC are complementary to each other. 

\acknowledgments{
I would like to thank S. Dawson for suggesting this project and thank Nathaniel Craig, Ian Lewis, and Marc Sher  
for helpful discussions.   This work is supported by the United States Department of Energy under Grant DE-AC02-98CH10886.
}

\vskip -.5in
\begin{table}[tp]
\renewcommand{\arraystretch}{1.4}
\caption{Measured Higgs Signal Strengths}
\centering
\begin{tabular}{|c|c|c|}
\hline
 Decay          & Production & Measured Signal Strength $R^{meas}$ \\ \hline
%\multirow{7}{*}
{$\gamma \gamma$} 
                & ggF        & $1.6^{+ 0.3 +0.3}_{-0.3-0.2}$, [ATLAS] \cite{atlas-13012}\\ 
                & VBF        & $1.7^{+ 0.8 +0.5}_{-0.8-0.4}$  [ATLAS]\cite{atlas-13012}\\ 
                & Vh        & $1.8^{+ 1.5 +0.3}_{-1.3-0.3}$  [ATLAS]\cite{atlas-13012}\\ 
                & inclusive  & $1.65^{+ 0.24 +0.25}_{-0.24-0.18}$   [ATLAS]\cite{atlas-13012}\\ 
                & ggF+tth    & $0.52 \pm 0.5$   [CMS]\cite{cms-13001}\\ 
                & VBF+Vh     & $1.48^{+1.24}_{-1.07}$  [CMS]\cite{cms-13001}\\ 
                & inclusive  & $0.78^{+0.28}_{-0.26}$      [CMS]\cite{cms-13001}\\
                & ggF        & $6.1^{+3.3}_{-3.2}$  [Tevatron]\cite{hcp:Enari}\\ \hline
%                
%\multirow{5}{*}
{$W W$}             
                & ggF        & $0.82 \pm 0.36$          [ATLAS] \cite{atlas-13030}\\ 
                & VBF+Vh     & $1.66 \pm 0.79$    [ATLAS]\cite{atlas-13030}\\ 
                & inclusive  & $1.01 \pm 0.31$   [ATLAS]\cite{atlas-13030}\\ 
                & ggF        & $0.76 \pm 0.21$        [CMS]\cite{cms-13003}\\ 
                & ggF        & $0.8^{+0.9}_{-0.8}$    [Tevatron]\cite{hcp:Enari}\\ \hline
%
%\multirow{2}{*}
{$ZZ$}              
                & ggF        & $1.8^{+0.8}_{-0.5}$   [ATLAS] \cite{atlas-13013}\\ 
                & VBF+Vh     & $1.2^{+3.8}_{-1.4}$   [ATLAS]\cite{atlas-13013}\\ 
                & inclusive  & $1.5 \pm 0.4$         [ATLAS]\cite{atlas-13013}\\ 
                & ggF        & $0.9^{+0.5}_{-0.4}$   [CMS] \cite{cms-13002}\\ 
                & VBF+Vh     & $1.0^{+2.4}_{-2.3}$   [CMS]\cite{cms-13002}\\ 
                & inclusive  & $0.91^{+0.30}_{-0.24}$ [CMS]\cite{cms-13002}\\ \hline
\end{tabular}

\vspace{-1ex}
%\caption{$*$ Only 8 TeV results are used in this channel.
%}
\label{tab:models1}
\end{table}

\begin{table}[tp]
\renewcommand{\arraystretch}{1.4}
\caption{Measured Higgs Signal Strengths}
\centering
\begin{tabular}{|c|c|c|}
\hline
 Decay          & Production & Measured Signal Strength $R^{meas}$ \\ \hline
%
%\multirow{5}{*}
{$b\bar{b}$}      
                & Vh         & $-0.4 \pm 1.0$    [ATLAS] \cite{atlas-170}\\ 
                & Vh         & $1.3^{+0.7}_{-0.6}$      [CMS]\cite{cms-044}\\ 
                & Vh         & $1.56^{+0.72}_{-0.73}$   [Tevatron]\cite{hcp:Enari}\\ \hline
%\multirow{8}{*}
{$\tau^+ \tau^-$}
                & ggF        & $2.4 \pm 1.5$          [ATLAS]\cite{atlas-160}\\ 
                & VBF        & $-0.4 \pm 1.5$         [ATLAS]\cite{atlas-160}\\ 
                & inclusive  & $0.8 \pm 0.7$          [ATLAS]\cite{atlas-170}\\ 
                & ggF        & $0.73 \pm 0.50$        [CMS]\cite{cms-13004}\\ 
                & VBF        & $1.37^{+0.56}_{-0.58}$  [CMS]\cite{cms-13004}\\ 
                & Vh         & $0.75^{+1.44}_{-1.40}$    [CMS]\cite{cms-13004}\\ 
                & inclusive  & $1.1 \pm 0.4$        [CMS]\cite{cms-13004}\\ 
                & ggF        & $2.1^{+2.2}_{-1.9}$    [Tevatron]\cite{hcp:Enari}\\ \hline                 
\end{tabular}

\vspace{-1ex}
\label{tab:models2}
\end{table}

\end{document}